\documentclass[sigplan,10pt]{acmart}
\renewcommand\footnotetextcopyrightpermission[1]{}

\setcopyright{none}
\usepackage{tikz}
\usepackage{amsmath}

\usepackage{filecontents}
\usepackage{booktabs}
\usepackage{makecell}
\usepackage{diagbox}
\usepackage{graphicx}
\usepackage{units}
\usepackage{blindtext}
\usepackage{xcolor,xspace}
\usepackage{ulem}
\usepackage{amsmath}
\usepackage{paralist}
\usepackage{mdframed}
\usepackage{pict2e}
\usepackage{mdframed}
\usepackage{xcolor}
\usepackage{threeparttable}
\usepackage{array}
\usepackage{pifont}
\usepackage{balance}
\usepackage{verbatim}
\usepackage{array}
\usepackage{booktabs}
\usepackage{float}
\usepackage{listings}
\usepackage{wrapfig}
\usepackage{tikz}
\usepackage{fancyhdr}
\usepackage{url}
\usepackage{framed}
\usepackage{pgfgantt}
\usepackage{xcolor}
\usepackage{listings}
\usepackage{caption}
\usepackage{soul}

\definecolor{asmblue}{RGB}{0,70,150}
\definecolor{asmred}{RGB}{160,40,40}
\definecolor{asmgray}{RGB}{100,100,100}
\definecolor{asmgreen}{RGB}{0,120,70}

\lstdefinelanguage{ARMasm}{
    morekeywords={
        mov,movz,movk,msr,mrs,bl,b,br,blr,ret,
        ldr,str,add,sub,and,orr,eor,cmp,tst,isb,
        cbz,cbnz,b.eq,b.ne,b.lt,b.gt,b.le,b.ge,bti
    },
    sensitive=false,
    morecomment=[l]{//},
    morecomment=[l]{;},
}

\usepackage{xcolor}

\lstdefinestyle{asmstyle}{
    language=ARMasm,
    basicstyle=\ttfamily\small,
    keywordstyle=\color{asmblue}\bfseries,
    commentstyle=\color{asmgray}\itshape,
    stringstyle=\color{asmgreen},
    identifierstyle=\color{black},
    columns=fullflexible,
    keepspaces=true,
    frame=single,
    framesep=5pt,
    breaklines=true,
    captionpos=t
}
\newcommand{\acro}{\textsc{ARA}\xspace}

\newcommand{\adv}{\textsc{Adv}\xspace}

\newcommand\ignore[1]{}
\newcommand\todo[1]{\textcolor{red}{TODO: #1}}

\newcommand{\inlinecode}[1]{\colorbox{lightgray!30}{\texttt{#1}}}

\begin{document}
%
\title{Extending Layered Attestation into ARM Application Run-Times}

\author{Antonio Joia Neto}
\email{antonio.joianeto@uzh.ch}
\affiliation{%
  \institution{University of Zurich}
  \country{}
}

\author{Bilel Sefsaf}
\email{bilel.sefsaf@uzh.ch}
\affiliation{%
  \institution{University of Zurich}
  \country{}
}

\author{Norrathep Rattanavipanon}
\email{norrathep.r@phuket.psu.ac.th}
\affiliation{%
  \institution{Prince of Songkla University,\\University of Zurich}
  \country{}
}
\author{Carlos Chinea Perez}
\email{carlos.chinea.perez@huawei.com}
\affiliation{%
  \institution{Huawei Technologies}
  \country{}
}

\author{Hans Liljestrand}
\email{hans.liljestrand2@huawei.com}
\affiliation{%
  \institution{Huawei Technologies}
  \country{}
}

\author{Jan-Erik Ekberg}
\email{jan.erik.ekberg@huawei.com}
\affiliation{%
  \institution{Huawei Technologies}
  \country{}
}

\author{Ivan De Oliveira Nunes}
\email{ivan.deoliveiranunes@uzh.ch}
\affiliation{%
  \institution{University of Zurich}
  \country{}
}


\begin{abstract}
Modern mobile and edge platforms provide hardware-backed attestation of application and platform integrity at load-time. However, they offer little support once application execution begins. We present ARA, an \underline{A}pplication-\underline{R}untime \underline{A}nchor for ARM that extends this load-time chain of trust into execution. ARA bootstraps from the underlying ARM platform Root of Trust (RoT) through a measured loading chain and establishes an attested protected run-time within the application address space using recently introduced ARM Permission Overlay Extensions (POE). The resulting run-time domain can host programmable RoTs for Measurement and Storage (RTM/RTS), enabling application run-time integrity verification, attestation evidence generation, and protected use of secrets while retaining efficient access to application run-time state. As exemplary uses, we instantiate ARA for protected media decryption and application control flow measurement. Our prototype and evaluation show that ARA supports frequent run-time RoT-backed operations at near in-process cost, with roughly 10× and 50× lower overhead than out-of-process equivalents. This offers a practical bridge from layered attestation to application run-time trusted computing on ARM.
\end{abstract}

\maketitle
\makeatletter
\fancypagestyle{reviewstyle}{
  \fancyhf{}
  \renewcommand{\headrulewidth}{0pt}
  \renewcommand{\footrulewidth}{0pt}

  \fancyhead[L]{\ACM@linecountL}
  \fancyhead[R]{\ACM@linecountR}

  \fancyfoot[C]{\thepage}
}
\makeatother

\pagestyle{reviewstyle}

%

\section{Introduction}\label{sec:intro}

Trusted computing architectures rely on Roots of Trust (RoTs) to establish confidence in system integrity and protect security-critical assets such as cryptographic keys, measurements, and security policies. While modern mobile and edge platforms increasingly deploy trusted computing services~\cite{apple_attest,android_integrity_1,android_integrity_2,android_integrity_3,android_integrity_4}, these mechanisms remain largely confined to layered attestation that ends at application load-time verification, essentially proving to a remote verifier (e.g., a remote application server) that the right application binary was launched on top of the right operating system. Once a user-space application begins executing, little trustworthy infrastructure remains available for measuring run-time application behavior or protecting security-critical state within the application itself. We argue that this limitation stems from a key challenge in establishing an application-centric RoT: \textit{how to simultaneously isolate a trusted (in-)application run-time substrate and connect this isolated substrate with the underlying hardware-backed RoT guarantees?}

Conventional attempts to achieve this goal typically rely on separate processes, privileged software (e.g., within the kernel), or  Trusted Execution Environments (TEEs)~\cite{azab2014hypervision,ge2014sprobes,payne2008lares,azab2009hima,costan2016intel,arm_cca}. While some of these approaches provide adequate isolation, run-time trusted computing operations must repeatedly cross protection boundaries to observe or process application state. This requires context switches, inter-process communication (IPC), or privilege level/TEE transitions. The resulting cost, thus, becomes prohibitive for run-time security services that interact with the application at fine granularity, e.g., execution/continuous integrity monitoring. On the other hand, software-only techniques can avoid some boundary-crossing costs, but instead rely on extensive software instrumentation, incurring significant run-time overheads to the application itself~\cite{tan2017principles,lefeuvre2025sok,SFI_fence}. These limitations call for run-time trusted computing approaches that both (1) remain rooted in platform hardware anchors and (2) avoid crossing traditional/costly isolation boundaries.

We also observe that, for the first time, the required hardware support for both hardware-backed layered attestation (currently built upon ARM TrustZone) and application-level isolation co-exist in the same architecture (ARM v8.9), with the recent introduction of ARM's Permission Overlay Extensions (POE) \cite{POE} (details in Sec.~\ref{sec:bg}). 
In particular, we argue that POE changes the design space for trusted computing on ARM by enabling strong isolation within a single process without requiring costly process, kernel, enclave, or hypervisor transitions, for application measurement and state-based computation. Building on this observation, we compose POE with layered attestation to design, implement, and evaluate Application Run-time Anchor (\acro). \acro is hardware-isolated and attested, executes within the application's address space (user space), and can be instantiated as run-time trusted computing services for application measurement and storage.

While in-process memory access control features similar to POE have been explored in x86 architectures, in particular based on Intel MPK~\cite{hodor,erim,secpoline} (see Section~\ref{sec:rw}), x86 lacks the hardware support for layered attestation, which at present is implemented on ARM devices.
We believe the composition of the two features to be a timely direction that enables several potential security applications. Trusted computing services are increasingly deployed to protect user-space applications on ARM-based edge devices (as exemplified by Android Play Integrity and Apple's App Attest system discussed in Section~\ref{sec:app_attestation}), and POE is a very recent technology that enables a new design space to further secure the application run-times. 

Our evaluation shows that \acro 
provides a practical and cost-effective foundation for run-time measurement, verification, and protection of user-space applications. In more detail, \acro benefits from three properties: (1) POE permissions are checked in parallel with Memory Management Unit (MMU) checks, and permissions can be reconfigured with one register write operation in user-space; (2) POE can be leveraged to make \acro code and data inaccessible (execution is only permitted via controlled invocation points) to application code; and (3) POE can be composed into layered attestation to ensure that the configuration of isolated trusted substrate in user space is verifiable.

The asymmetry in in-process memory access control provided by POE allows us to instantiate \acro with RoT functions without requiring IPC, kernel mediation, or similar. In other words, \acro is isolated from, but shares the same virtual address space as the application. This makes security-relevant computation, e.g., logging relevant control flow transfers and system calls (syscalls) for attestation/auditing, enforcing execution integrity against control flow graphs, confidentially computing on secrets, and signing attestation reports, as cheap as normal in-process computation. Simultaneously, \acro's integration with layered attestation provides hardware-backed remote verifiability.

\textbf{Intended Contributions.} In sum, our intended contributions are:
\begin{itemize}
\item {\bf \acro Architecture:} an in-process run-time trust anchor that extends ARM platform attestation into user-space execution using POE.
\item {\bf Secure Construction:} a bootstrapping and launch-binding protocol plus controlled invocation/syscall mediation that preserves the ARA boundary even after the host application is compromised.
\item {\bf RTS/RTM Instantiations:} we provide concrete RTS and RTM services, demonstrated through protected media decryption and in-process control-flow attestation.
\item {\bf Prototype and Evaluation:} we provide an open-source \acro prototype implementation and respective evaluation showing roughly an order-of-magnitude performance improvement on the RTS workload and roughly 30–55× improvement over process isolation for the high-frequency RTM workload.
\end{itemize}

\section{Background}\label{sec:bg}

\subsection{Roots of Trust (RoTs)}

An RoT is a minimal Trusted Computing Base (TCB) used to establish the security foundation of a computing system. Because higher-level security mechanisms depend on the integrity of it, RoTs are typically designed to be verifiable and isolated from less trusted software layers. Common examples include secure elements, Trusted Platform Modules (TPMs), immutable boot ROMs, and hardware security modules (HSMs). In modern systems, RoTs enable functions such as secure boot, cryptographic key management, and platform remote integrity verification (a.k.a., remote attestation). RoTs are generally categorized as RoTs for measurement and storage\footnote{Some of the literature distinguishes a third category, namely RoT for Reporting (RTR), used when the RTM result is to be signed to generate an externally verifiable report that can be checked by a remote verifier, e.g., in remote attestation protocols. In this paper, we deliberately couple RTR as a sub-function of RTM.}.

A {\bf Root of Trust for Measurement (RTM)} focuses on system integrity by securely measuring software (either statically, i.e., at load-time or run-time/execution properties). Examples of measured properties include load-time code integrity and run-time integrity (e.g., control and data flow).

A {\bf Root of Trust for Storage (RTS)} provides secure protection for sensitive cryptographic material and other security-critical data. Unlike an RTM, which focuses on integrity measurement, the RTS ensures the confidentiality and controlled usage of secrets such as encryption keys, certificates, authentication credentials, and attestation keys. In many architectures, cryptographic operations involving protected keys are performed entirely within the RTS boundary, preventing key material from ever being exposed in plaintext to untrusted software.

\acro establishes a run-time RoT within a user-space process. Section~\ref{sec:evaluation} presents two case studies demonstrating how \acro can be instantiated as RTM and RTS for user-space process execution.

\subsection{Layered Attestation and Application Integrity}\label{sec:app_attestation}

Modern mobile platforms (such as iOS and Android) provide hardware-backed RTMs for attestation services that allow application servers to verify the integrity of both application code and aspects of the underlying execution environment at or near load-time. Two prominent examples are Apple’s App Attest~\cite{apple_attest} and Google's Play Integrity API~\cite{android_integrity_1}. These systems aim to reduce the risk of tampering, repackaging, instrumentation, and kernel-level compromise by binding cryptographic assertions to device and application state.

At a high level, these services rely on ARM architectural support for trusted hardware~\cite{trustzone,apple_enclave,titan}. When an application is launched, the platform generates a cryptographic key pair that is bound to the application identity and the device's hardware. The private key never leaves the trusted hardware boundary and is used to sign attestation statements that can be verified by a remote backend using the corresponding public key and platform-issued certificate chain.

\begin{figure}
  \centering
    \includegraphics[width=\columnwidth]{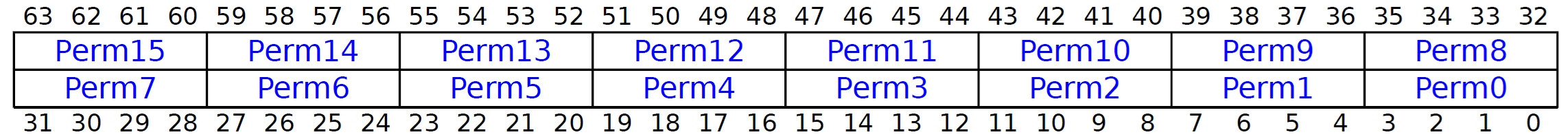}
    \caption{\texttt{POR\_EL0} structure. Used to assign R/W/X permissions for domains based on their IDs}\label{fig:porel0}
\end{figure}

During application launch or shortly thereafter, the application server can request (via a challenge-response protocol) an attestation token that reflects both the integrity of the application code and evidence about the device state. Evidence includes whether the OS has been modified, whether boot integrity checks have passed, and whether known debugging or hooking frameworks are present. These include kernel-level integrity indicators derived from verified boot and system security configuration. The resulting attestation object is sent to a backend application server, which validates the cryptographic signature and evaluates the included integrity claims against a policy. If the attestation passes, the server can treat the client as running in a ``known-good'' state and engage in sensitive operations.

While widely used~\cite{android_integrity_2}, current integrity attestation platforms only provide load-time measurements and leave run-time integrity unchecked. Intuitively (and as shown empirically in Section~\ref{sec:evaluation}), this limitation stems from the high overhead that run-time application RoT functions would have under typical isolation models (e.g., process-based), which must rely on costly context switches and IPC. \acro addresses this gap by establishing an overlayed protected domain within the process itself and connecting it into the layered attestation trust chain.

\subsection{Permission Overlay Extension (POE)}
\label{subsec:background-poe}

A recent hardware feature, introduced in the ARMv8.9 architectures, is Permission Overlay Extension (POE)~\cite{POE}. It enforces access control policies at runtime with minimal overhead, allowing software to dynamically restrict memory access among domains within the same process without requiring kernel mediation.

POE acts as an additional access control layer on top of conventional virtual memory protections. Access permissions are enforced at the granularity of virtual memory pages using Page Table Entries (PTEs). POE associates memory pages with overlay identifiers (IDs). These IDs encode logical protection domains. During execution, the processor maintains a permission register -- namely, \texttt{POR\_EL0} -- that specifies the currently allowed access rights, Read, Write, eXecute (R/W/X), for each overlay domain ID.

As depicted in Figure~\ref{fig:porel0}, \texttt{POR\_EL0} stores 16 permissions (4 bits for each domain over its 64-bit size). However, with the standard 64-bit page-table entries, the domain ID stored in the PTE is 3 bits wide, allowing only 8 different domain IDs and, therefore, only \texttt{Perm0}--\texttt{Perm7} to be used. The remaining entries, \texttt{Perm8}--\texttt{Perm15}, are reserved in this configuration~\cite{POE}. In practice, this means that 8 domain permissions can be configured simultaneously. The position of the bits in the layout of \texttt{POR\_EL0} defines the domain ID for the enforcement of that permission. For example, permissions for the domain of ID 7 (\texttt{Perm7}) are stored in bits 28 to 31.

\begin{figure}
  \centering
    \includegraphics[width=0.9\columnwidth]{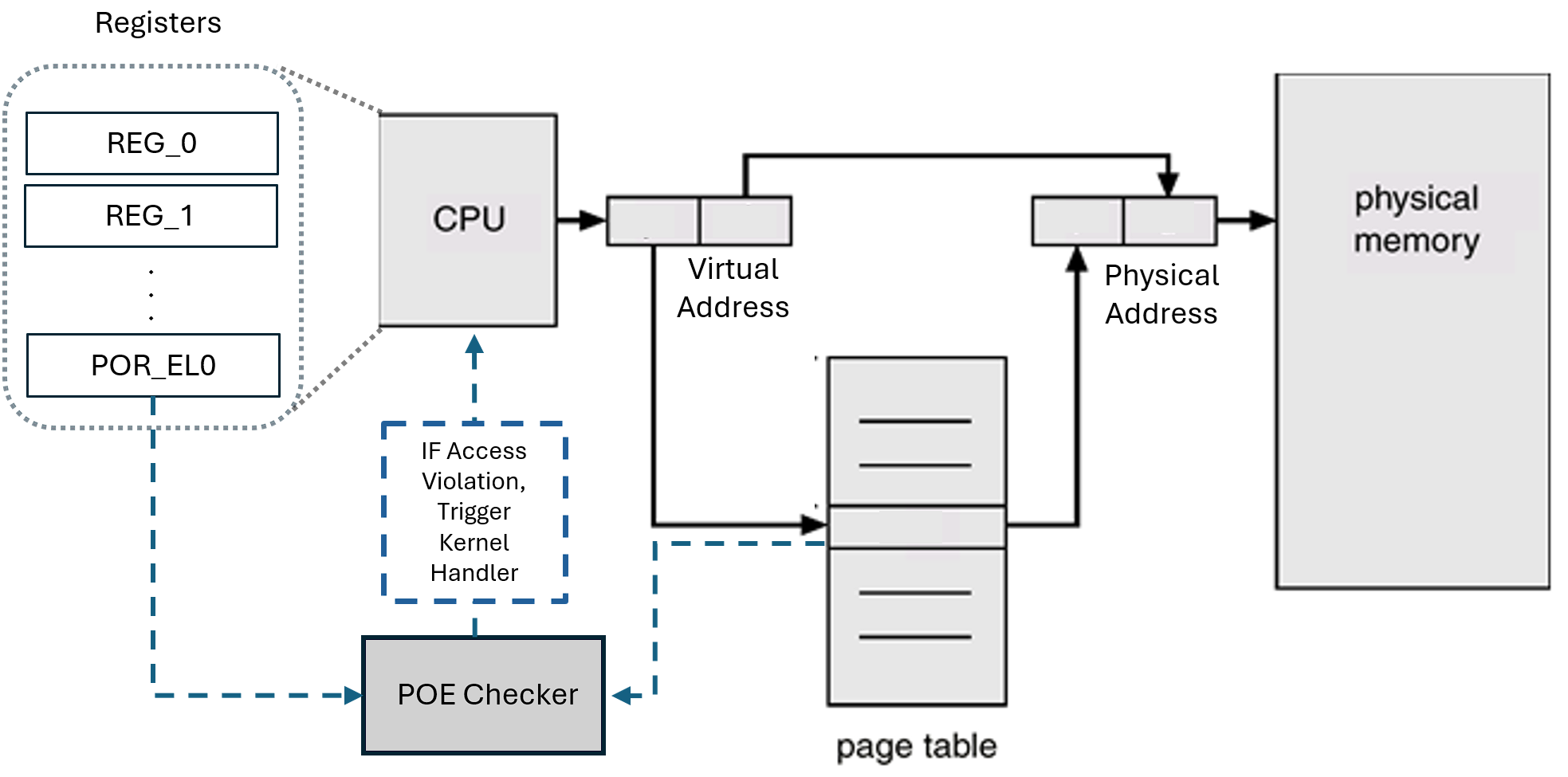}
    \caption{POE Access Check. Domain Identifiers in PTEs are checked against $POR\_EL0$ at runtime.}\label{fig:poe_arch}
\end{figure}

When a memory page is created, its PTE is annotated with the protection domain to which the page belongs. POE encodes this domain using four PTE bits that identify the page’s overlay index. On each memory access, the processor enforces both the regular page-table permissions and the active \texttt{POR\_EL0} configuration for the overlay index encoded in the target page’s PTE. The effective permission is therefore the intersection of the conventional MMU permission and the POE permission selected by that page’s domain ID.
These checks occur in hardware, yielding minimal runtime overhead. This is illustrated in Figure~\ref{fig:poe_arch}. As a result, multiple isolated domains can share the same virtual address space.

\textbf{Fast Domain Switching.} A key feature of POE is the ability to reconfigure domain permissions through a simple and lightweight register update to \texttt{POR\_EL0} in user-space, rather than requiring kernel mediation.
Overwriting \texttt{POR\_EL0} when switching execution contexts within a process redefines the access permissions of each domain in one instruction. The instruction \texttt{msr} (Move to System Register) must always be used to set \texttt{POR\_EL0}. Therefore, application code can be statically inspected at loading time to ensure that unprivileged executable domains do not contain instructions with structure such as \texttt{msr POR\_EL0 <X>}, that would overwrite \texttt{POR\_EL0} illegally. Because \texttt{POR\_EL0} updates occur entirely in user-space, domain transitions are significantly faster, enabling otherwise impractical isolation models.


Since Linux 6.12, arm64 provides upstream support for the POE with the \texttt{pkey\_*} system-call family and assign pagess to a specific domain using \texttt{pkey\_mprotect(addr, len, prot, pkey)}. In this model, Linux exposes this functionality through the standard pkey abstraction.

\section{Assumptions \& Goals}\label{sec:overview}

\subsection{System Model}

We assume that the underlying device system features an ARM architecture equipped POE (see POE details in Section~\ref{subsec:background-poe}). Furthermore, we assume that the system features a code and platform integrity measurement system (as discussed in Section~\ref{sec:app_attestation}) such as Apple App Attest, Google Play Integrity, or similar, along with the hardware support required by these systems (e.g., ARM TrustZone). This model is consistent with typical client-side edge devices running ARM Cortex-A processors, such as mobile phones, laptops, IoT devices, etc. We further assume that the device's layered attestation (a.k.a. platform integrity) mechanism is properly implemented by the vendor. 

\subsection{Adversary (\adv) Model}
 
 We assume a user space \adv capable of compromising application level components within a process, e.g., via memory corruption to attempt to launch attacks (such as code injection, control, and data flow attacks), while the underlying hardware (including POE enforcement and MMU) is trusted. 
 Once checked by the underlying platform integrity measurement system (e.g., Google Play Integrity, Apple App Attest, etc. -- recall Section~\ref{sec:app_attestation}), the kernel can be trusted. This is not an additional assumption: it is always required by process-based user space isolation, regardless of \acro. The attacker may attempt to read or tamper with sensitive data, escalate privileges within the process (i.e., into \acro), or interfere with security services implemented by \acro{}, including invoking illegal instructions, abusing \acro calls/entry points, or attempting to reconfigure domain permissions. However, \adv cannot directly bypass hardware-enforced permission checks. Microarchitectural side-channels~\cite{meltdown,spectre,sok_side_channel1,sok_side_channel2} are treated as orthogonal since they are mitigated by the underlying CPU micro-architecture~\cite{cache}. 

\subsection{Goals}\label{subsec:goal}


    \noindent [\textbf{G1}] \textbf{Run-time trust retention.} Once the platform integrity mechanism has accepted the expected ARA-enabled application configuration and ARA has completed its trusted initialization, subsequent ARA-generated security outputs must remain attributable to that protected run-time instance despite \adv compromise of the application (App). 
    
    \noindent [\textbf{G2}] \textbf{Integrity of run-time measurements.}
    For an RTM, the run-time evidence produced by \acro must faithfully reflect the run-time state processed by the RTM logic.
    In particular, \adv controlling App must not be able to tamper with the measurement logic, corrupt the measurement state, or forge run-time evidence that would be accepted as legitimately produced by \acro.

    \noindent [\textbf{G3}] \textbf{Confidentiality and integrity of \acro data.} Sensitive data managed by \acro must remain protected from \adv. 
    In particular, \adv must not be able to directly read/write security-critical data protected within \acro.

    \noindent [\textbf{G4}] \textbf{Execution integrity of \acro.} \acro must execute only intended functionality. Specifically, \adv may interact with \acro only through valid service calls and must not be able to alter their execution (e.g., via unintended control flows that bypass validation logic or cause \acro to return through an unauthorized exit.






\subsection{Non-Goals}\label{sec:non-goals}

\acro is designed to establish a run-time anchor within an already load-time attested user space process. Its objective is to enable trusted measurement and security services that are isolated from, yet able to efficiently interact with, an untrusted application. Consequently, the following goals fall outside the scope of this work.

\textit{Protection against an attested yet compromised platform.}
\acro assumes the integrity of the underlying platform, including the hardware, firmware, boot chain, attested operating system kernel, and the platform attestation mechanism used to bootstrap trust (i.e., attest the kernel and loaded application). This is in line with current platform integrity services discussed in Section~\ref{sec:app_attestation}.

\textit{Replacing TEEs.}
\acro is not intended to replace TEEs such as TrustZone~\cite{trustzone} and SGX~\cite{sgx}. TEEs protect applications from a potentially untrusted platform by providing isolated execution environments and hardware-backed confidentiality guarantees. In contrast, \acro builds upon layered attestation systems (see Section~\ref{sec:app_attestation}) commonly found in mobile and edge devices to protect trusted run-time services from a compromised application while remaining within the application's address space. The two approaches are complementary and address different threat models.

\textit{General-purpose compartmentalization.}
The goal of \acro is not to provide a generic multi-compartment isolation framework. Rather, it establishes a small privileged attested application substrate that functions as a run-time RoT for the host process. While POE could support broader compartmentalization designs, exploring scalable multi-domain isolation policies is outside the scope of this work.

\section{\acro Overview}

We present \acro rationale as a common substrate for instantiations such as RTMs and RTSs for application run-times.

\subsection{\acro Load-Time + Layered Attestation}

\begin{figure}
  \centering
    \includegraphics[width=0.9\linewidth]{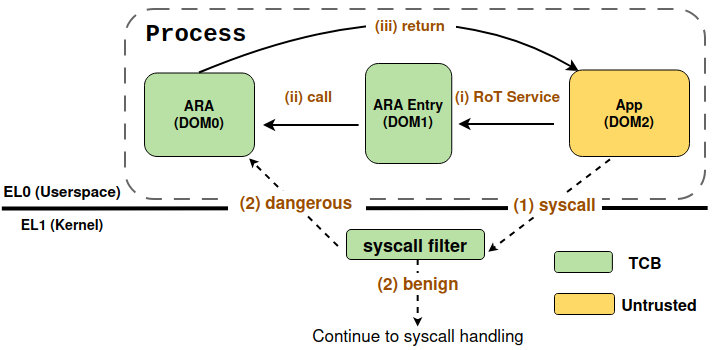}
    \caption{Architectural Overview of \acro}\label{fig:overview}
\end{figure}

\acro load-time aims to establish and attest a user space process with the configuration shown in Figure~\ref{fig:overview}. The process must contain three isolated domains: \acro domain, controlled \acro invocation domain, and the (untrusted) application, denoted App.
Further, a syscall filter~\cite{endokernel} is also bootstrapped at process load-time and enforced by the kernel during run-time.

Across all domains, \acro maintains default MMU page access permissions: code pages are executable and non-writable; immutable pages are configured as read-only, and mutable data pages are readable and writable.

The first domain to execute is always \acro for {\bf secure initialization of run-time configurations and App}.
At that stage, all virtual memory objects are also assigned to this privileged \acro domain.
Since \acro code is the first to run, it initially executes with all (RWX) permissions to all domains.
Next, it configures POE to establish the required memory access controls that isolates the three domains appropriately before other domains can execute.

\acro load-time deliberately makes POE configurations a part of \acro domain's code, which is the first to execute.
This design choice is aimed at ensuring that \acro initialization behavior (including the setup of POE configurations, App binary static inspection, and syscall filter setup) can be captured when utilizing the underlying layered attestation system to attest the program load-time configuration. In sum, by design, this leads to the attestation of the code of \acro domain proving that the process configuration shown in Figure~\ref{fig:overview} was performed. This is because the configuration is a static part of the binary loaded onto the \acro domain.

In turn, this allows \acro secure bootstrapping to extend the layered chain of trust. The platform integrity service measures the process binary, i.e., the \acro domain code. \acro then inspects and loads the App binary, performs POE access control configurations accordingly, before transferring control to App.
This yields the chain

\begin{equation*}
\textbf{Platform RTM} \rightarrow \textbf{\acro loader} \rightarrow \textbf{App}    
\end{equation*}

where each stage verifies the next at load time. Since this design requires only load-time measurement of the process itself, it fits nicely onto what platform integrity attestation services already support.

\subsection{POE Access Control at Run-Time}\label{sec:overview_runtime}

As mentioned above, POE is also configured by \acro before App can execute.
Specifically, the process address space is partitioned into three POE-protected domains: DOM0 for \acro itself (trusted code), DOM1 for controlling \acro  invocation, and DOM2 for App (untrusted code).

Table~\ref{table:domain-config} summarizes the POE configurations performed by \acro. Each row specifies the active \texttt{POR\_EL0} access-control configuration applied when code executes in a given domain (the subject domain), while each column indicates the access permissions to another domain (the object domain). For example, when App code executes, \texttt{POR\_EL0} is configured according to the DOM2 row. This configuration prevents App from directly accessing \acro while still allowing execution to enter DOM1, which serves as \acro's legal entry point.

\begin{table}[t]
\centering
\caption{User-space process POE configuration with \acro.}
\label{table:domain-config}
\footnotesize
\setlength{\tabcolsep}{3.5pt}
\begin{tabular}{lccc}
\toprule
\makecell[l]{\textbf{Subject} $\downarrow$ \\ \textbf{Object} $\rightarrow$}
&
\makecell{\textbf{\acro}\\\textbf{DOM0}}
&
\makecell{\textbf{\acro Entry}\\\textbf{DOM1}}
&
\makecell{\textbf{App}\\\textbf{DOM2}} \\
\midrule
\textbf{DOM0}          & \texttt{RWX} & \texttt{RWX} & \texttt{RWX} \\
\textbf{DOM1 (entry)}  & --           & \texttt{X}   & \texttt{RWX} \\
\textbf{DOM1 (exit)}   & \texttt{RWX} & \texttt{RWX} & \texttt{RWX} \\
\textbf{DOM2}          & --           & \texttt{X}   & \texttt{RWX} \\
\bottomrule
\end{tabular}
\end{table}

Second, \acro registers a seccomp-BPF filter~\cite{seccomp} with the kernel.
The idea is similar to that used in~\cite{endokernel} but adapted for \acro's requirements.
The filter intercepts syscalls issued by App and redirects potentially dangerous ones to \acro DOM0 for arbitration at run-time.
Dangerous syscalls include those modifying memory mappings or indirectly violating POE permissions.
We discuss them in more detail in Section~\ref{sec:syscall-trap}.

After completing the setup, \acro switches to the DOM2 transfering control to App's first instruction (e.g., its \texttt{main} function) to start App execution.

App executes normally until it issues a syscall or \acro call, shown by transitions (1) and (i) in Figure~\ref{fig:overview}, respectively.

A syscall is first checked by the seccomp-BPF filter installed at load time.
Inoffensive syscalls proceed through normally;
whereas dangerous system calls are redirected to \acro together with their arguments, allowing \acro to perform additional checks before deciding whether or not to perform this syscall.

App may also invoke \acro services at any point during run-time.
As App executes under the DOM2 configuration, it cannot directly jump into \acro in DOM0 (per Table~\ref{table:domain-config}).
The only permitted path into \acro is through the dedicated \acro contolled invocation in DOM1, for which DOM2 has execute permission.

This controlled invocation path is shown by the solid line in Figure~\ref{fig:overview}: App must always invoke an \acro secure services through their legal entry point.
The entry point mediates the transition into \acro by executing under the DOM1 POE initial configuration (DOM1 entry). DOM1 then elevates privilege (DOM1 exit) and performs the required context switch into DOM0. DOM0 executes the requested operation, restores the DOM2 POE configuration, and returns the result to App.
We detail this controlled execution flow in Section~\ref{sec:controlled-invocation}.

\subsection{\acro Instantiations}

Beyond the basic initial configurations, the \acro domain also implement configurable (and load-time attestable) trusted computing services that can be securely called as described in Section~\ref{sec:overview_runtime}.

~\\\noindent \textbf{Secret Provisioning}

\acro further supports a provisioning mechanism for receiving secrets from an external trusted party (e.g., an application server or another process) which we refer to as the \emph{Provisioner}.
This mechanism combines \acro with the underlying hardware-backed integrity attestation platform to let the provisioner verify that the process was initialized with the expected code and system configuration.

\acro can obtain platform-integrity evidence for the initialized process and associate this evidence with an \acro-controlled provisioning endpoint.
\acro can then forward the evidence to the \emph{Provisioner} (a.k.a. \emph{Verifier}), which verifies the evidence against the expected load-time state, including the expected \acro code, App code, and provisioning endpoint.
Upon successful verification, the \emph{Provisioner} encrypts a secret to the \acro-controlled in-process endpoint and sends it back to \acro. Recall from Section 2 that integrity attestation platforms can be used to provision a public key to the attested end-point.
As a result, the secret is provisioned into \acro in DOM0 and never exposed directly to App in DOM2.
We note that the exact mechanism by which the provisioning endpoint is bound to platform-integrity evidence depends on the underlying platform. Details are discussed in Section~\ref{sec:platform-binding}.

\vspace{-1mm}
~\\\noindent \textbf{RTS Instantiations}

From the provisioning mechanism for delivering secrets into the protected \acro domain discussed above, what remains for instantiating \acro as an RTS is to implement the logic for protecting and using these secrets for confidentiality purposes.
In particular, RTS services must ensure that provisioned secrets remain within DOM0 and that outputs returned to App through \acro calls (mediated by DOM1 controlled invocation) reveal only the intended result, not the underlying secret or unintended functions of the secret.
This allows App to use provisioned secrets indirectly through \acro in-process calls, without directly accessing them.
For example, an RTS service may use a provisioned master secret to decrypt application-specific secrets or derive session keys, while keeping the master secret inaccessible to App.
In Section~\ref{subsec:case_study_rts}, we present a concrete case study in which \acro-based RTS uses a provisioned master secret to support protected decryption for video streaming.

~\\\noindent \textbf{RTM Instantiations}

Instantiating \acro as an RTM requires defining what runtime state should be measured and how that state is delivered to \acro.
This typically requires modifying (e.g, via instrumentation) App to collect and forward the relevant runtime state to \acro through the \acro calls.
Once control reaches \acro, the RTM-specific measurement logic measures/extends the received runtime state and then resumes App execution.

In a typical end-to-end flow, the application server first uses the provisioning mechanism at launch to verify the process's load-time state and establish a signing key with \acro.
For RTM, this verification also confirms that the expected App instrumentation was present at load time (in addition to \acro trusted domain).

Later, during process execution, the server can request runtime evidence from \acro.
\acro generates a report from the runtime events collected through the deployed instrumentation, signs it using the provisioned signing key, and sends the signed evidence to the server.
We present a concrete RTM case study in Section~\ref{subsec:eval-rtm}.

\section{Design Details}\label{sec:design}





\subsection{Bootstrapping and measuring \acro TCB within a process} 
\label{sec:binding iprot}

\acro configuration must be securely bootstrapped and measured within the process before App starts and within the same process.
An \acro user-space loader is the first to execute on creation of an \acro process. At launch, the standard platform integrity mechanism can be used to measure the loader as part of the initially created process.
The loader (which is also measured in the previous step) is preconfigured with a known-good hash of the code expected to run within \acro trusted DOM0. Before loading any code, the loader verifies the hash of the code to be loaded against the known-good value. As a result, trust in the measured loader extends to the code loaded within \acro, effectively creating a layered measurement chain for the process, similar to the trust chains used in secure boot architectures~\cite{arbaugh1997secure}.

The loader first loads an \acro service library (namely \texttt{libara.so}) and hands control to its constructor.
Since \texttt{libara.so} has been verified, its constructor is trusted to complete the remaining initialization.
This trusted constructor performs the following: (i) verify the loaded App and the \acro entry point against expected hashes, (ii) scan App's executable pages and rewrite illegal \texttt{POR\_EL0} instructions (Section~\ref{subsec:binary-analysis}), (iii) install the seccomp-BPF filter (Section~\ref{sec:syscall-trap}), and map/tag the POE domains for the \acro entry point (DOM1) and App (DOM2).
After these steps, \acro transfers control to App.
From this point on, App execution is unprivileged, and DOM0 is reachable only through controlled invocation points (Section~\ref{sec:controlled-invocation}) or through fault/exception handling (Section~\ref{sec:syscall-trap}).

The resulting trust chain begins with the platform-integrity mechanism authenticating the expected ARA-enabled application and platform state. The ARA-aware loader then verifies the ARA run-time components, including \texttt{libara.so}, the ARA entry point, and App, before untrusted execution begins. In this way, trust established at the platform level is extended to the specific in-process ARA configuration.

During provisioning, this chain allows the Provisioner to accept an ARA-controlled endpoint only when it is associated with the expected authenticated application configuration and the ARA components verified during initialization. ARA therefore does not require the underlying platform-integrity service to directly measure individual ARA components or to be modified to support ARA. Instead, component-level verification is performed by the trusted ARA loading chain.


\subsection{Load-Time Binary Analysis of App.}\label{subsec:binary-analysis}

At load time, \acro statically inspects all executable App code and removes every instruction that can write to \texttt{POR\_EL0}. This approach is necessary because \texttt{POR\_EL0} is writable from userspace (EL0), allowing an untrusted App to modify its own permissions by executing \inlinecode{msr POR\_EL0, Xn} (\texttt{Xn} denotes an arbitrary source register).
Since AArch64 instructions are fixed-width and 4-byte aligned, the \acro constructor walks each word-aligned address in every executable region, matches the fixed opcode and system-register encoding of \inlinecode{msr POR\_EL0, Xn} while masking the source-register field, and replaces each match with a NOP. This simplifies instruction filtering compared to x86, where instructions are not aligned.

To prevent such instructions from being introduced after the initial scan, \acro also mediates all transitions that create or modify executable memory, including side-loaded code and JIT-generated code. Any request from the untrusted App (DOM2) to make memory executable, such as a JIT promoting an \texttt{RW} page to \texttt{RX}, is trapped and serviced only after \acro maps the page read-only, scans it, and patches any prohibited instruction. Thus, no DOM2-executable page can contain an instruction that writes to \texttt{POR\_EL0}.


\subsection{Syscall Filtering and Mediation}
\label{sec:syscall-trap}

The POE domain configuration prevents App from directly accessing \acro's protected memory, However, App may still invoke kernel interfaces that indirectly violate the isolation boundary. For example, a compromised App could ask the kernel to change page permissions, retag PTEs, or copy data from/to DOM0 memory. \acro adopts the syscall-mediation approach and classification of security-sensitive kernel interfaces established by  prior work ~\cite{secpoline,schrammel2022jenny,cerberus,endokernel}.

During initialization, \acro installs a seccomp-BPF filter that intercepts security-sensitive syscalls originating outside \acro's code region. These include syscalls that can modify address space or protection state, affect POE domain assignments, or allow the kernel to access protected DOM0/DOM1 memory on App's behalf.  Intercepted App syscalls are redirected to \acro, which validates their arguments before allowing or rejecting the operation. Signal delivery and return are similarly mediated, following the general approach of prior MPK-based isolation systems, to prevent App from regaining control with an unauthorized \texttt{POR\_EL0} configuration.

To avoid the additional seccomp trap and signal-handling overhead for frequently invoked syscalls, \acro also exposes a syscall service through its controlled invocation interface. App passes the syscall number and arguments directly to \acro, which applies the same validation policy and, if permitted, issues the syscall from DOM0 on App's behalf. Thus, the fast path changes only how the syscall request reaches \acro, it does not weaken the syscall-mediation policy.

\subsection{Controlled Invocation}
\label{sec:controlled-invocation}


All RoT services are implemented inside \texttt{libara.so} and execute in DOM0. However, App cannot invoke these services directly since DOM2 has no execute permission over DOM0. Instead, App must enter through the \acro entry point, whose code is placed in DOM1. We expose the \acro entry point through the interface \texttt{secure\_call(FUNCTION\_ID fid, void *arg)} that can be accessed by App, where \texttt{fid} identifies the requested service with input arguments in \texttt{arg}.

As shown in Listing~\ref{lst:por-el0-write}, the entry point saves the App context, switches \texttt{POR\_EL0} to the DOM0 configuration, and executes a memory/instruction barrier (\texttt{isb}) to make the new permissions effective. 
Afterwards, the control is transferred to the trusted dispatcher, which switches the stack pointer to DOM0 stack, validates the requested service and its arguments, and invokes the corresponding function.

\begin{lstlisting}[style=asmstyle, caption={\acro controlled entry point.}, label={lst:por-el0-write},
basicstyle=\scriptsize\ttfamily,
% numbers=left,
numberstyle=\tiny,
stepnumber=0.8,
numbersep=10pt]
secure_call:            // DOM1 entry gate 
  ...                   // save App execution context
  mov  x9, DOM0_CONFIG  
  msr  POR_EL0, x9      // enable DOM0 permissions
  isb                   
  b    ara_secure_call  // enter trusted DOM0 dispatcher
\end{lstlisting}

On transferring back from DOM0 to DOM2, \acro clears its execution state, restores the DOM2 POE configuration, and resumes the saved App context. The \texttt{POR\_EL0} permission switch is placed at the DOM0--DOM2 page boundary so that, after DOM0 execute permission is removed, the next instruction is fetched from a page executable in DOM2.

\subsection{Platform Integrity Service Integration: Google Play Integrity and Apple App Attest}
\label{sec:platform-binding}


\acro leverages the underlying platform-integrity mechanism to authenticate the expected application and platform state and to bind application-provided data to the resulting integrity evidence. This binding is used to associate a provisioning endpoint, identified by a public key $pk^{D0}$, with the ARA-enabled application. This endpoint is later used by the Provisioner to deliver secrets encrypted under $pk^{D0}$ to \acro. The exact mechanism used to establish this binding depends on the platform. Here, we discuss how it can be realized on commonly deployed platforms: Android and iOS.

\textbf{Google Play Integrity.}
During \acro initialization, \acro generates an endpoint key pair $(pk^{D0}, sk^{D0})$ and includes $pk^{D0}$ in the \texttt{requestHash} field of a Play Integrity request.
The resulting integrity token is forwarded to the server, which asks Google's Play Integrity backend to verify the token and return the verdict.
The server accepts only if: (i) the returned \texttt{requestHash} matches $pk^{D0}$; (ii) the application-integrity verdict is satisfied (i.e., \texttt{appRecognitionVerdict = PLAY\_RECOGNIZED} and the signing-certificate digest matches with that of the expected application); and (iii) the platform-integrity verdict is satisfied.
On success, the server provisions the secret by encrypting it under $pk^{D0}$.

\textbf{Apple App Attest.}
During initialization, \acro requests an App Attest key using
\texttt{DCAppAttestService.}\texttt{generateKey}, which returns an ID for a key managed by Apple's App Attest service.
\acro then calls \texttt{DCAppAttestService.attestKey} with this key ID and sends the resulting attestation object to the application server.
The server validates the attestation object against Apple's trust chain, checks that the attested key is bound to the expected App ID, and extracts the attested public key $pk^{D0}$.
The server accepts only if the attestation object is valid for the expected App ID.
On success, the server encrypts the secret using $pk^{D0}$.

\section{Security Analysis}\label{sec:security-analysis}

We first argue that invariants \textbf{IV1}-\textbf{IV7} (defined below) hold during process execution.
Then, we show them sufficient to achieve security goals \textbf{G1-G4} defined in Section~\ref{subsec:goal}.


    \noindent [\textbf{IV1}] \textbf{No POE misconfiguration by App.}
    App-controlled executable pages in DOM2 cannot contain instructions that modify \texttt{POR\_EL0}. This follows from the load-time binary analysis (Section~\ref{subsec:binary-analysis}) and syscall mediation preventing unchecked executable-code injection at run-time (Section~\ref{sec:syscall-trap}). Writes contained in the fixed DOM1 transition sequence are handled separately by \textbf{IV3}.

    \noindent [\textbf{IV2}] \textbf{App always resumes with DOM2 privilege.}
    Whenever control is transferred to App, the active POE access configuration corresponds to DOM2.
    This holds at load time, when \acro installs the DOM2 configuration before starting App, and at run-time, when \acro restores the DOM2 configuration before every return from DOM0 (Section~\ref{sec:controlled-invocation}) or when delivering a mediated signal/exception to App (Section~\ref{sec:syscall-trap}).

    \noindent [\textbf{IV3}] \textbf{No direct App $\rightarrow$ \acro control transfer.}
    App cannot directly execute DOM0 code. Any successful transfer from App into DOM0 must pass through the DOM1 entry point, and from DOM 1 only the fixed DOM0 entry instruction. Arbitrary entry within DOM1 cannot result in App-controlled execution with DOM0 permissions (Section~\ref{sec:controlled-invocation}).

     \noindent [\textbf{IV4}] \textbf{Mediated dangerous syscalls.}
    Any DOM2-issued syscall that could access DOM0 (with R/W/X) or DOM1 (R/W) memory or modify MMU/POE permissions or metadata is denied by \acro.
    This is realized using the seccomp-BPF filter described in Section~\ref{sec:syscall-trap}.

    \noindent [\textbf{IV5}] \textbf{Mediated asynchronous control flows.}
    All user-space signal and exception deliveries are first routed to \acro.
    \acro validates the event, installs the DOM2 POE configuration before exposing it to App, and mediates signal return so App cannot restore a non-DOM2 configuration.
    This is enforced by \acro's signal-delivery and signal-return mediation, as described in Section~\ref{sec:syscall-trap}.

    \noindent [\textbf{IV6}] \textbf{Trusted service correctness.}
    \acro and RoT service handlers are trusted and assumed to be implemented correctly.
    In particular, they must be memory-safe, validate App-provided inputs before use, and return only outputs permitted by the service policy.
    This is a standard trusted computing/RoT assumption (Section~\ref{sec:overview}).

    \noindent [\textbf{IV7}] \textbf{Load-time binding.}
    The platform-integrity mechanism authenticates the expected \acro-enabled application and platform configuration. Before untrusted App execution begins, the \acro-aware loading chain verifies the \acro run-time components, including \texttt{libara.so}, the \acro entry point, and App. Together, these steps bind the provisioning/signing endpoint to the expected initialized \acro process.

\subsection{G1: Run-time Trust Retention}

We first establish that the trust established during \acro initialization remains valid after App becomes adversarial. \textbf{IV7} ensures that the initialized \acro instance, entry point, and App correspond to the expected load-time configuration. After initialization, \textbf{IV1–IV5} preserve the \acro boundary. App cannot enable unauthorized POE permissions, execute DOM0 directly, indirectly modify \acro through dangerous kernel interfaces, or use asynchronous control flow to resume with elevated permissions. \textbf{IV6} also requires individual RoT services to protect their own trusted state and validate untrusted input.

\subsection{G2: Integrity of Run-time Measurements}

For an RTM instantiation, the measurement logic and measurement state reside in DOM0.
\textbf{IV7} rules out the goal's violation at load-time, e.g., replacing App, removing instrumentation, or using an unexpected \acro instance. 

At this point, the remote server knows that the expected App code (with instrumentation, e.g., RoT-relevant calls, see Section~\ref{subsec:eval-rtm} for examples) is loaded.
What remains to be shown is that with instrumentation in place \adv cannot tamper with measurement at run-time.
This violation can be done by:
(1) directly modifying the RTM state or logic in DOM0;
(2) forcing \acro code to execute along an unintended path;
(3) leverage syscalls to modify measurement state or permissions;
(4) abuse signals/exceptions to deviate control flows bypassing instrumentation; or
(5) trick an RoT service into corrupting state or producing forged evidence.

Case (1) is blocked by \textbf{IV1} and \textbf{IV2} since these two invariants together imply that at any point, App only has access to DOM2 configuration and thus cannot directly modify the RTM state/logic in DOM0.
We argue how (2) is prevented as part of {\bf G4} arguement below.
Protection from (3), (4) and (5) directly follow from \textbf{IV4}, \textbf{IV5} and \textbf{IV6}, respectively.
Therefore, \acro design satisfies \textbf{G2}.

\subsection{G3: Confidentiality and Integrity of \acro Sensitive Data}

For an RTS instanciation, sensitive data is stored inside DOM0.
By compromising App, \adv can attempt to access or modify such data in three ways: (1) directly from App in user space, (2) indirectly through kernel syscalls, or (3) through an authorized \acro service call.
(1) is prevented by \textbf{IV1} and \textbf{IV2}; without a valid DOM0 configuration, App in DOM2 has no means to access data inside DOM0.
(2) is prevented by \textbf{IV4}, which denies or mediates DOM2 syscalls that could access or modify DOM0 memory.
Finally, (3) is limited by the execution integrity of \acro (see below in {\bf G4} argument) and by \textbf{IV6}, where a valid call to \acro services returns only outputs permitted by their service policy, and therefore do not reveal the underlying protected secrets.
Thus, \acro design satisfies \textbf{G3}.

\subsection{G4: Execution Integrity of \acro}\label{subsec:sec-analysis-g3}

G4 requires that App cannot force \acro to execute along unauthorized paths.
At run-time, \adv can attempt this in three ways:
(1) directly transferring control from App to DOM0 code;
(2) leveraging syscalls to modify code, permissions, or execution state in DOM0; or
(3) abusing asynchronous control flow, such as signals or exceptions.


(1) is prevented by \textbf{IV3}. Any transfer from App to DOM0 must pass through the DOM1 entry point, whose only transfer into DOM0 targets the fixed dispatcher \texttt{ara\_secure\_call}. Thus, App cannot choose an arbitrary DOM0 instruction as the entry target or directly enter the middle of an \acro function.

Although App may branch to an arbitrary instruction within the DOM1 entry point, this cannot result in App-controlled execution with DOM0 permissions. Before the \texttt{msr} that updates \texttt{POR\_EL0}, the DOM2 configuration remains active. If App enters after the \texttt{msr}, the DOM2 configuration likewise remains active, and the subsequent transfer to DOM0 faults because DOM0 is not executable.

App may also enter directly at the \texttt{msr} with an attacker-controlled value in the source register. However, after this instruction, the only instructions before entering DOM0 are the \texttt{isb} and a direct branch to the fixed \texttt{ara\_secure\_call} dispatcher. Therefore, if the resulting configuration does not permit DOM0 execution, the transfer faults; if it does, control necessarily switches to the trusted DOM0 dispatcher before any App-controlled instruction can execute. Asynchronous signals or exceptions occurring during this transition are covered by \textbf{IV5}, which ensures that control is returned to App only under the DOM2 configuration.

(2) is prevented by \textbf{IV4}, which denies or mediates DOM2 syscalls that could tamper with \acro code, permissions, or execution state.
(3) is prevented by \textbf{IV5}, since signals and exceptions are routed to \acro first and are delivered to App only after validation under the DOM2 configuration.
Therefore, App cannot hijack the \acro entry point, enter the middle of an \acro function, skip validation logic, or force unauthorized partial execution of \acro code, satisfying \textbf{G4}.

\ignore{
We argue security through invariants established before the application runs and inductively preserved thereafter.  We reason in three phases: load time, run-time, and post-execution.

\textbf{Load-Time.} At load time, trust is anchored in the underlying platform integrity measurement system (as described in Section~\ref{sec:binding ara}), which verifies the correctness of the chain $\textbf{trusted loader} \rightarrow \texttt{libara.so} \rightarrow \textbf{App}$, before App execution. It also performs POE tagging and installs the seccomp filter before App execution. Consequently, these load-time steps cannot be skipped. Conditioned on this, four invariants are established. \textbf{(L1)} Initialization completes in an adversary-free window, since DOM2 gains control only after \texttt{POR\_EL0} has been narrowed. \textbf{(L2)} the DOM2 configuration enforces no access to DOM0, and execute-only access to the DOM1 entry. \textbf{(L3)} Every \texttt{msr POR\_EL0} on a DOM2-executable page is rewritten to \texttt{NOP}. \textbf{(L4)} The seccomp filter is installed under \texttt{NO\_NEW\_PRIVS}, rendering it irremovable, and is origin-bound to the DOM0 range.

Argue in terms of invariants.\\

- hijecking the entry point

1- What holds before an \acro application runs.\\

2- What holds while an \acro application runs (and why this is true due to 1).\\

3- What holds after an \acro application dies (and why this is true due to 1 and 2).\\

Refer to system assumptions, threat model, and design while explaining why 1, 2, and 3 hold.\\
}

\section{Prototype and Evaluation}\label{sec:evaluation}

Since POE is not yet available in commercial devices, we construct two complementary prototypes: (1) a QEMU instance implementing full POE support on a modified ARM Cortex-A76 processor model, and (2) a physical ARM Cortex-A76 processor. In case (2), we replace each \texttt{msr} instruction that writes to \texttt{POR\_EL0} with an \texttt{msr} instruction that writes to \texttt{NZCV} instead, while keeping required memory barriers. Because the written \texttt{NZCV} value is never used during execution, this substitution preserves program behavior while accounting for the cost of the register update on App$\leftrightarrow$\acro transitions.


\subsection{RTS Case Study: Isolated Media Decryption}\label{subsec:case_study_rts}

This case study instantiates \acro{} as an RTS, with a workload inspired by Digital Rights Management (DRM) deployments, e.g., Google's Widevine L3 \cite{google_widevine}: a software mechanism used for DRM in browsers and mobile devices. In Widevine L3, an application receives encrypted media streams over a standard network connection and delegates the decryption of the media chunks to a dedicated component. If this component is appropriately isolated (a separate process in Widevine L3 path), this prevents a compromised host application from accessing the cryptographic keys or the media in plaintext. Our evaluation isolates the core decryption component and quantifies the performance trade-offs across three different configurations:

\begin{itemize}
    \item \textbf{Baseline (No Isolation):} Executes all components within the same process without isolation.
    \item \textbf{Process-Based Isolation:} decryption is performed in a separate process. In this case, we implement three variations of common IPC methods: Shared Memory (SHM), Unix domain Sockets, and named Pipes (FIFO).
    \item \textbf{\acro{}-Based Isolation:} both components exist within the same process, decryption function and associated cryptographic key is isolated within \acro{}.
\end{itemize}

Process-Based Isolation also represents a lower bound on the cost of using the ARM Cortex-A TEE (TrustZone) instead, because TEE calls involve a superset of the context switches required for normal IPC~\cite{pinto2019demystifying}.

\textbf{Implementation Details.}
To realize this case study, we implement an HTTP-fed media decryption pipeline using the \texttt{gpac/libgpac} framework~\cite{gpac}. The input media is protected with AES-based authenticated encryption.
Our experiment assumes that the required decryption key is provisioned to the \acro-protected component once by the application server upon successful load-time attestation\footnote{This is in addition to standard user authentication and any other typical server-side checks.} performed by the underlying platform integrity measurement system (recall Section~\ref{sec:app_attestation}). Our results thus focus on the process run-time. In each instance we measure the decryption time for the full MP4 media workload.

We evaluate three workload variations, summarized in Table~\ref{tab:drm_videos}. These test cases vary media resolution and thus sizes (2~GB, 4~GB, and 8~GB) to evaluate whether isolation overhead scales with data volume under sustained load. 
Each media file is 10 minutes long and uses one-second movie fragments.
Media bytes are delivered to the decryption component in 64~KiB logical chunks. For the shared memory configuration, we use a 4~MiB shared-memory ring buffer.

\begin{table}
\centering
\caption{MP4 videos used in the RTS case study.}
\label{tab:drm_videos}
\small
\setlength{\tabcolsep}{3pt}
\begin{tabular}{lccccc}
\toprule
Video & Dur. & Resolution & FPS & Size & Bitrate \\
\midrule
1080p & 600 s & 1920$\times$1080 & 60 & 2.00 GB & 26.71 Mbps \\
4K    & 600 s & 3840$\times$2160 & 60 & 4.00 GB & 53.37 Mbps \\
4K    & 600 s & 3840$\times$2160 & 60 & 8.01 GB & 106.74 Mbps \\
\bottomrule
\end{tabular}
\end{table}

We pin all configurations to a single Cortex-A76 core to accurately measure the overhead of the isolation boundaries under identical hardware conditions. Using a single core prevents parallelism from masking run-time overhead across different configurations.

\begin{figure}[t]
  \centering
    \includegraphics[width=0.85\linewidth]{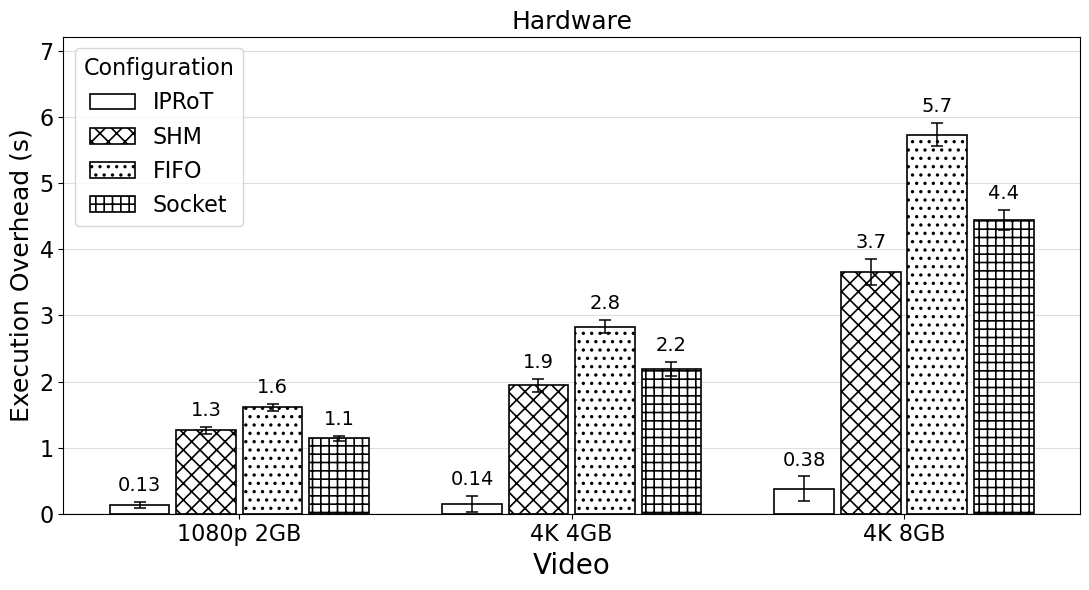}
    \caption{Overheads of RTS case on physical hardware.}\label{fig:drm_hw}
\end{figure}

\begin{figure}
  \centering
    \includegraphics[width=0.85\linewidth]{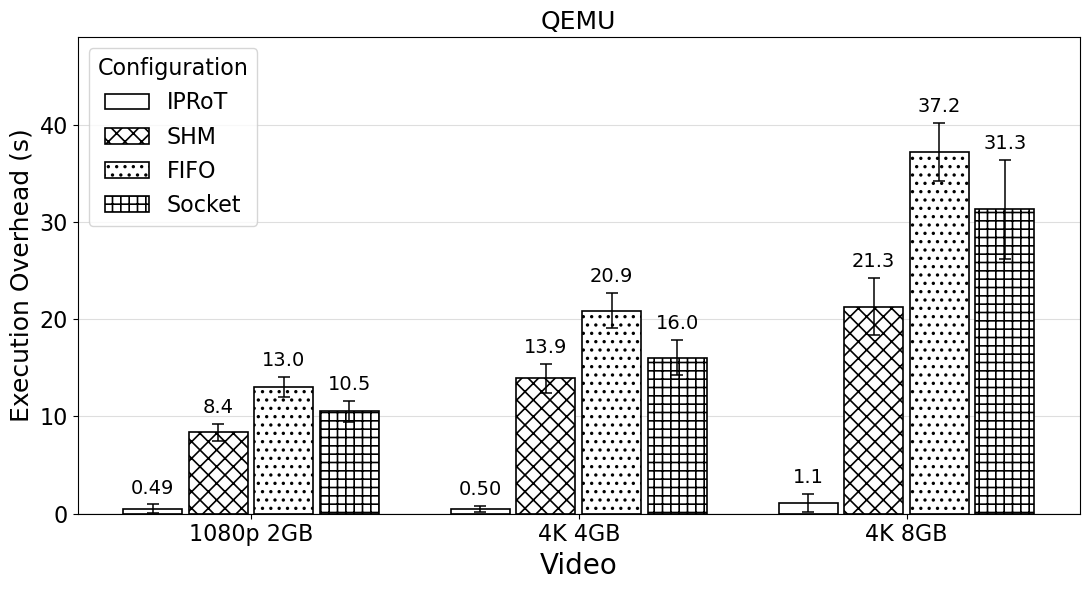}
    \caption{Overheads of RTS case on QEMU.}\label{fig:drm_qemu}
\end{figure}

\textbf{Results.}
Figures~\ref{fig:drm_hw} and \ref{fig:drm_qemu} show the isolation overhead (as compared to the baseline case without isolation) for the three test cases running on native-hardware
proxy and QEMU. All results are averaged across 50 repetitions, with error bars denoting 95\% confidence intervals.

On the physical Cortex-A76 (Figure~\ref{fig:drm_hw}) \acro incurs less overhead than process-based isolation for all IPC mechanisms. For the 2GB video, \acro{} adds on average 0.13 seconds of overhead, whereas Sockets, SHM, and FIFO add 8$\times$ to 12$\times$ more isolation overhead than \acro{}. This disparity stems from system time spent on context switches and kernel memory reads. Processing the 2GB video forces SHM and FIFO into 120,000 and 123,000 context switches, and Unix sockets into 41,000, as measured using GNU \texttt{/usr/bin/time -v}. While SHM overhead is dominated by these context switches, sockets and FIFO incur further latency from system calls reading and writing system memory. \acro{} bypasses these IPC yields by maintaining an in-process security boundary, executing the workload with 220 context switches, due to standard Linux periodic execution (e.g., scheduling). This trend scales as expected with larger workloads. For the 8 GB video, \acro{} adds 0.38 seconds of overhead, while SHM, FIFO, and Unix sockets add 9.7$\times$, 15$\times$, and 11.6$\times$ more overhead than \acro{}, respectively.

We additionally report results using our POE-enabled QEMU implementation. They show the same qualitative trend as the physical hardware evaluation, with \acro{} overhead remaining roughly 10x lower than process-based isolation with all IPC-based alternatives. ignoreThe wider confidence intervals reflect the non-determinism inherent to emulated environments. Despite this variance, the separation between \acro{} and the IPC baselines remains clearly distinguishable.

\begin{table*}[]
\centering
\caption{Run-time overhead of RTM case study on SPEC2026 benchmarks \ignore{\todo{check omnet  CFA INstructions value}}}
\label{tab:RTM Results}
\resizebox{0.9\linewidth}{!}{%
\begin{tabular}{c|llll|ll|llcc}
 & \multicolumn{4}{c|}{\textbf{Application Information}} & \multicolumn{2}{c|}{\textbf{\acro}} & \multicolumn{4}{c}{\textbf{IPC}} \\ \cline{2-11} 
{\textbf{Application}} & \multicolumn{1}{c}{\begin{tabular}[c]{@{}c@{}}Conditional \\ Branches\end{tabular}} & \multicolumn{1}{c}{\begin{tabular}[c]{@{}c@{}}Indirect \\ Branch\end{tabular}} & \multicolumn{1}{c}{Return} & \multicolumn{1}{c|}{\begin{tabular}[c]{@{}c@{}}CFA  \\ Instructions\end{tabular}} & \multicolumn{1}{c}{\begin{tabular}[c]{@{}c@{}}QEMU\\ run-time \\ (ms)\end{tabular}} & \multicolumn{1}{c|}{\begin{tabular}[c]{@{}c@{}} ARM Cortex \\ A76\\ (ms)\end{tabular}} & \multicolumn{1}{c}{\begin{tabular}[c]{@{}c@{}}QEMU\\ run-time \\ (ms)\end{tabular}} & \multicolumn{1}{c}{\begin{tabular}[c]{@{}c@{}}QEMU\\ run-time \\ overhead\end{tabular}} & \begin{tabular}[c]{@{}c@{}}ARM Cortex \\A76 \\ (ms)\end{tabular} & \begin{tabular}[c]{@{}c@{}}ARM Cortex \\ A76\\ Overhead\end{tabular} \\ \hline
cpython & 83080 & 1579 & 13522 & 98181 & 8247 $\pm$ 14 & 1085 $\pm$ 12 & 455375 $\pm$ 311 & {+ 5421\%} & 55513 $\pm$ 37 & {+ 5016\%} \\
flightdm & 20142 & 1986 & 27916 & 50044 & 32214 $\pm$ 80 & 4916 $\pm$ 31 & 1525675 $\pm$ 693 & {+ 4636\%} & 192320 $\pm$ 37 & {+ 3812\%} \\
lbm & 98 & 0 & 23 & 121 & 6871 $\pm$ 89 & 892 $\pm$ 25 & 233476 $\pm$ 559 & {+ 3303\%} & 27501 $\pm$ 27 & {+ 2983\%} \\
ntest & 2447 & 54 & 3959 & 6460 & 4522 $\pm$ 46 & 655 $\pm$ 18 & 215288 $\pm$ 470 & {+ 4761\%} & 27995 $\pm$ 11 & {+ 4174\%} \\
omnetpp & 53550 & 20764 & 147364 & 325898 & 868 $\pm$ 42 & 109 $\pm$ 22 & 46560 $\pm$ 244 & {+ 5264\%} & 5678 $\pm$ 13 & {+ 5109\%} \\
sqlite & 16750 & 204 & 2173 & 19127 & 2929 $\pm$ 66 & 499 $\pm$ 13 & 13778$\pm$ 158 & {+ 4603\%} & 2350 $\pm$ 29 & {+ 4609\%} \\
stockfish & 4057 & 35 & 5001 & 9093 & 66921 $\pm$ 59 & 19369 $\pm$ 28 & 3663907 $\pm$ 879 & {+ 5374\%} & 937198 $\pm$ 61 & {+ 4738\%} \\
zstd & 7893 & 28 & 1622 & 9543 & 4942 $\pm$ 7 & 659 $\pm$  11 & 156333 $\pm$ 517 & {+ 3063\%} & 24321 $\pm$16 & {+ 3590\%} \\ \hline
\end{tabular}%
}
\end{table*}

\subsection{RTM Case Study: Execution Integrity.}\label{subsec:eval-rtm}

Contrary to RTS, the RTM case study aims to evaluate a security service that imposes high RoT call frequencies. We therefore instantiate \acro as a Control-Flow Attestation (CFA)~\cite{sok} RoT.
CFA records security-relevant control-flow events during execution and produces evidence that can be inspected by a remote verifier for run-time auditing. As noted in the recent systematization of~~\cite{sok}, prior CFA largely focused on bare-metal software (usually embedded micro-controllers) or required privilege-level and/or process-level isolation and associated costs. This case study shows how ARA can be used to avoid both limitations.

To implement CFA, App is instrumented at compile time via an LLVM~\cite{llvm} pass. This passes non-deterministic control flow transfers with RoT calls to log their destinations into \acro protected memory.
Each address is appended to a buffer of size 1 MB. Upon filling the buffer, the RTM aggregates the recorded transfers into a hash chain:
$ H \leftarrow \mathrm{Hash}(H || buffer )$. 
Once a partial CFA report is requested from a Verifier
(in this case, \emph{Provisioner} that attested the application load-time configuration), the security requirement is that a compromised App must not be able to forge or tamper with the recorded sequence of addresses, ensuring that they faithfully reflect App run-time behavior.

The cost of CFA is determined by the number of control flow destinations
appended to the isolated log (within the RTM). Cost of logging, in turn,  depends on how the log is isolated from App. Under the same CFA policy we evaluate two isolation models:

\begin{itemize}
  \item \textbf{\acro}: The log resides in DOM0. Each control-flow transfer is recorded via a \texttt{secure\_call} that saves the address in a secure buffer within DOM0.
  \item \textbf{IPC Socket}: The log is isolated within a separate process. Since a log stored in App's writable memory could be tampered with or deleted by a compromised App prior to commitment.
\end{itemize}

\textbf{Applications and Setup.}
We evaluate with applications from the SPEC CPU 2026 suite~\cite{madhav2026spec}: \texttt{cpython},
\texttt{flightdm}, \texttt{lbm}, \texttt{ntest}, \texttt{omnetpp},
 \texttt{sqlite}, \texttt{stockfish}, and \texttt{zstd}.
These cover a wide range of control flow event densities, from interpreter- and search-heavy workloads with dense indirect control flow transfers (\texttt{cpython}, \texttt{stockfish}, \texttt{omnetpp}) to compute-bound numeric kernels whose hot loops contain only sparse instrumented transfers (\texttt{lbm}, \texttt{zstd}). All configurations use the reference
inputs and are pinned to a single Cortex-A76 core, on both the QEMU model and
the physical hardware. We report the average of 20 runs, which were sufficient to obtain small confidence intervals. The values reported are the average run-time $\pm$ 95\% confidence intervals.

\textbf{Results.}
Table~\ref{tab:RTM Results} reports run-time results per application.
\acro is 30 to 55 times faster relatively to process isolation. IPC requires a synchronous hand-off per edge, the producer must block until the logger commits the edge, since one left in application, writable memory could be rewritten or dropped first, incurring a context switch and kernel-mediated copy per transfer. \acro commits each control flow transition through \texttt{secure\_call} in-process,  whose overhead is closer to that of a plain function call. The variance in the results are related with the control-flow density of each application, where  bigger overheads are dominated by interpreter and search-heavy workloads (\texttt{cpython}, \texttt{stockfish}) while it is amortized in sparse numeric kernels (\texttt{lbm}).

\section{Discussion and Limitations}\label{sec:limit}









\textbf{Multi-Threading.}
Our prototype assumes serialized ARA execution / single-threaded App execution. To support multi-threaded applications, \acro run-time must be made thread-safe by protecting shared \acro state with appropriate synchronization primitives. At the architectural level, POE permissions are maintained as part of each thread's state and are automatically saved and restored by the kernel during context switches. As a result, each thread retains its own POE permission state when scheduled.

\textbf{Multiple \acro Instances in One Procces.} Our discussion thus far considers one \acro instance per process. Since POE supports eight simultaneously configurable domains, the remaining domains could host additional mutually distrusting domains. For example, an RTS holding cryptographic keys could exist in one domain, and an RTM holding measurement state in another, without mutual trust. This would require extending the access-control matrix of Table~\ref{table:domain-config} and allocating services (syscall mediation, secret provisioning, etc.) as appropriate for the particular setting of interest.

\textbf{JIT and Dynamic Code.} 
\acro design supports JIT and dynamically loaded code by mediating syscalls that makes memory executable (Section~\ref{subsec:binary-analysis}). Any DOM2 request to promote a page to executable is trapped, after which \acro maps the page read-only, scans and patches any illegal \texttt{POR\_EL0} writes, and only then retags it as executable DOM2 memory. This preserves invariant \textbf{IV1} at the cost of one run-time pass on the binary being dynamically loaded.



\section{Related Work}\label{sec:rw}

In practice, most current edge systems rely on classic inter-process isolation or TEEs. Regarding research, POE is very recent, and we are aware of only one preprint -- NanoZone~\cite{nanozone} -- that focuses on it. NanoZone extends support beyond the eight domains available in POE by re-labeling PTE domain IDs (a limitation noted in Sec.~\ref{sec:rw}), trading off performance for scalability (i.e., increased number of domains). Outside the POE-specific space, our work relates to several domains.

\textbf{User-Space Trusted Computing.} Current mainstream approaches to verify user-level process properties (notably Linux Integrity Measurement Architecture~\cite{sailer2004design,linux_ima,bohling2020subverting}, Android's Play Integrity API~\cite{android_integrity_1,android_integrity_2,android_integrity_3,android_integrity_4}, and Apple App Attest~\cite{apple_attest}) are restricted to the static state of the process at loading time (essentially attest that the binary has not been illegally modified) without support for measurement of run-time state. This is largely due to the complexity of measuring run-time state from outside the process. Our work bridges this gap by enabling an in-process RoT for run-time verification and run-time security services. We emphasize that enclaved execution systems such as Intel SGX~\cite{costan2016intel,fei2021security,zheng2021survey}, SANCTUM~\cite{costan2016sanctum}, Keystone~\cite{lee2020keystone}, etc., do not fall under our definition of in-process RoTs. They establish confidential and remotely attestable code segments (enclaves) within processes based on hardware RoTs. While they support outsourced confidential computation, these are not privileged with respect to the process itself and cannot implement the same type of services that integrate with layered system-wide attestation systems such as the abovementioned (thus differing from our goal).

\textbf{Software-Based Approaches.}
Related to isolation, in-process software-based compartmentalization techniques aim to constrain the impact of software vulnerabilities by restricting how different parts of a program interact and by enforcing boundaries during execution. Memory safety approaches~\cite{memsafe1,memsafe2,memsafe3,memsafe4} attempt to eliminate vulnerabilities that could lead to control- or data-flow attacks and data corruption before deployment. This is commonly achieved either through programming languages that integrate memory safety by design, such as Rust~\cite{rust}, which uses static compile-time analysis, ownership, and borrowing mechanisms to ensure temporal and spatial safety, or through memory-safe dialects of existing languages, such as Checked C~\cite{checkedc}, which augments C with safer pointer and array types and introduces spatial safety checks during compilation and run-time. A complementary line of work focuses on hardening code against exploits with mechanisms such as CHERI~\cite{cheri} (which requires hardware support and typically recompilation), software fault isolation~\cite{SFI_original,webass1,webass2,webass3}, memory layout randomization~\cite{aslr}, and control-flow integrity (CFI)~\cite{nyman2020toward}. Unlike this work, these techniques focus on strengthening code rather than constructing isolated in-process RoTs.

\textbf{User-Space Memory Protection in x86.} Intel Memory Protection Keys (MPK)~\cite{mpk} provides in-process access-control mechanisms resembling POE in more complex x86 architectures. ERIM~\cite{erim}, Hodor~\cite{hodor}, and Secpoline~\cite{secpoline} isolate state in x86 using MPK. Other notable approaches on MPK include extending MPK’s limited domain count using virtualization~\cite{mpk,gu2022epk} (similar to NanoZone's case in POE); enforcing heap metadata \cite{blair2022mpkalloc} and cross-language\cite{kirth2022pkru} integrity; and intra-unikernel isolation between trusted and untrusted boundaries~\cite{sung2020intra}. To improve on MPK and similar technologies, approaches such as LatticeBox~\cite{latticebox} investigate efficient ways of representing domain IDs to make domain transitions faster and domain ID representations more efficient, allowing fewer bits to represent more domains within a PTE. Prior systems used MPK primarily for protecting libraries or compartments. Further, x86 deployments do not feature platform integrity services, e.g. Google Play Integrity, Apple App Attest (see Section~\ref{sec:app_attestation}). This results in an orthogonal design space and set of goals: (1) establishing in-process RTM and RTS building upon platform/loadtime trusted computing, and (2) enabling these services on (predominantly ARM-based) edge/client devices.

\textbf{HW/SW Co-Designs for Memory Access Control.} 
 CCTAG~\cite{cctag} proposes integrating multiple disjoint memory safety mechanisms into a unified system. More recently, Lattice-BOX~\cite{latticebox} introduced a scheme for representing memory permissions that improves both the expressiveness of memory domains and the efficiency of domain switching by reducing reliance on memory barriers on Intel MPK-like systems. These efforts complement our work but focus on different aspects of execution/metadata storage optimization.

\textbf{Qualitative Comparison.} 
Table~\ref{tab:architecture-comparison} offers a summarized qualitative comparison between \acro and related systems.
 While existing MPK-based systems offer in-process isolation, they lack architectural support for hardware-backed layered attestation of their in-process run-times. Meanwhile, enclaved execution systems such as Intel SGX are confidential but unprivileged with respect to their host application, being unable to securely monitor App execution.


\begin{table}[t]
    \centering
    \caption{Qualitative comparison with related approaches.}
    \label{tab:architecture-comparison}
    \footnotesize
    \setlength{\tabcolsep}{3pt}
    \begin{tabular}{lccc}
        \toprule
        \textbf{Approach} &
        \makecell{\textbf{In-process}\\\textbf{isolation}} &
        \makecell{\textbf{HW-backed}\\\textbf{trust chain}} &
        \makecell{\textbf{Direct App-state}\\\textbf{access}} \\
        \midrule
        Separate Process & No  & No  & No  \\
        MPK-based        & Yes & No  & Yes \\
        SGX/Enclave      & Yes & Yes & No  \\
        \textbf{ARA}     & \textbf{Yes} & \textbf{Yes} & \textbf{Yes} \\
        \bottomrule
    \end{tabular}
\end{table}

\section{Conclusion}\label{sec:conclusion}

We presented \acro, a mechanism for extending trust established by platform integrity mechanisms into applications. \acro design eliminates the need for costly process, kernel, enclave, or hypervisor transitions traditionally required for application-level trusted computing services, enabling practical run-time roots of trust for measurement and storage on ARM-based edge and mobile devices. Our case studies and evaluation demonstrate that \acro can support security services such as run-time integrity measurement and protected secret handling with substantially lower overhead than conventional isolation approaches. More broadly, our results suggest that memory-permission mechanisms can bridge the long-standing gap between load-time user-space attestation and efficient run-time verification.


\bibliographystyle{ACM-Reference-Format}
\bibliography{references}

\end{document}